\documentclass[sigconf]{cidr-2025}
\makeatletter
\def\@ACM@checkaffil{
    \if@ACM@countrypresent\else
        \ClassWarningNoLine{\@classname}{No country present for an affiliation}%
    \fi
}
\makeatother
\usepackage{graphicx} 
\usepackage{subfigure}
\usepackage{xspace}
\usepackage{pifont}
\usepackage{tabularx,ragged2e}

\newcolumntype{C}{>{\centering\arraybackslash}X}

\definecolor{darkgreen}{rgb}{0.0, 0.5, 0.0}
\definecolor{darkred}{rgb}{0.5, 0.0, 0.0}

\newcommand{\code}{\texttt}

\newcommand{\cmark}{{\color{darkgreen}\ding{51}}} 
\newcommand{\xmark}{{\color{darkred}\ding{55}}}   

\newcommand{\myparagraph}[1]{\vspace{1mm}\noindent\textbf{#1}}

\newcommand{\uqo}{UQO\xspace}
\newcommand{\uqostar}{UQO*\xspace}

\newcommand{\revision}[1]{{#1}}
\title{Towards Query Optimizer as a Service (QOaaS) in a Unified LakeHouse Ecosystem: Can One QO Rule Them All?}

\author{Rana Alotaibi$^{*}$$^{1}$, Yuanyuan Tian$^{2}$, Stefan Grafberger$^{*}$$^{3}$, Jes\'us Camacho-Rodr\'iguez$^{5}$, Nicolas Bruno$^{2}$, Brian Kroth$^{2}$, Sergiy Matusevych$^{2}$, Ashvin Agrawal$^{2}$, Mahesh Behera$^4$, Ashit Gosalia$^4$, \\Cesar Galindo-Legaria$^{5}$, Milind Joshi$^{5}$, Milan Potocnik$^{5}$, Beysim Sezgin$^{5}$, Xiaoyu Li$^{5}$, Carlo Curino$^{2}$}
\affiliation{
 \institution{\institution{$^1$SDAIA\ \ \ \ $^2$Gray Systems Lab, Microsoft \ \ \ \ $^3$BIFOLD \& TU Berlin \ \ \ \ $^4$Fabric Spark, Microsoft \ \ \ \ $^5$Fabric DW, Microsoft}}
}
\renewcommand{\shortauthors}{Rana Alotaibi et al.}
\begin{abstract}

Customer demand, regulatory pressure, and engineering efficiency are the driving forces behind the industry-wide trend of moving from siloed engines and services that are optimized in isolation to highly integrated solutions.
This is confirmed by the wide adoption of open formats, shared component libraries, and the meteoric success of integrated data lake experiences such as Microsoft Fabric.

In this paper, we study the implications of this trend to Query Optimizer (QO) and discuss our experience of building Calcite and extending Cascades into QO components of Microsoft SQL Server, Fabric Data Warehouse (DW), and SCOPE.
We weigh the pros and cons of a drastic change in direction: moving from bespoke QOs or library-sharing (\`a la Calcite) to rewriting the QO stack and fully embracing Query Optimizer as a Service (QOaaS).
We report on some early successes and stumbles as we explore these ideas with prototypes compatible with Fabric DW and Spark.
The benefits include centralized workload-level optimizations, multi-engine federation, and accelerated feature creation, but the challenges are equally daunting.
We plan to engage CIDR audience in a debate on this exciting topic.

\end{abstract}

\begin{document}

\maketitle

\section{Introduction}

After two decades of ``one size does not fit all'', we found ourselves with a collection of high-performance engines and services that customers had to string together to solve end-to-end application scenarios. The response, driven by customer demand and engineering efficiency, is a convergence of several specialized engines to: 1) operate out of a single copy of the data in the lake, 2) share compute resources, and 3) provide single sign-on and governance experience. Microsoft Fabric \cite{fabric} represents the earliest example of this trend. Further evidence that our industry is trying to move past siloed engines is the emergence of regulator-friendly open standards such as Parquet~\cite{parquet}, Arrow~\cite{arrow}, Substrait~\cite{substrait} and system-building libraries such as Calcite~\cite{begoli2018apache}, Orca~\cite{soliman2014orca}, Datafusion~\cite{datafusion} or Velox~\cite{pedreira2022velox}. 

\begin{figure}[bpht]
    \centering
    \vspace{-4mm}
        \includegraphics[width=0.4\textwidth]{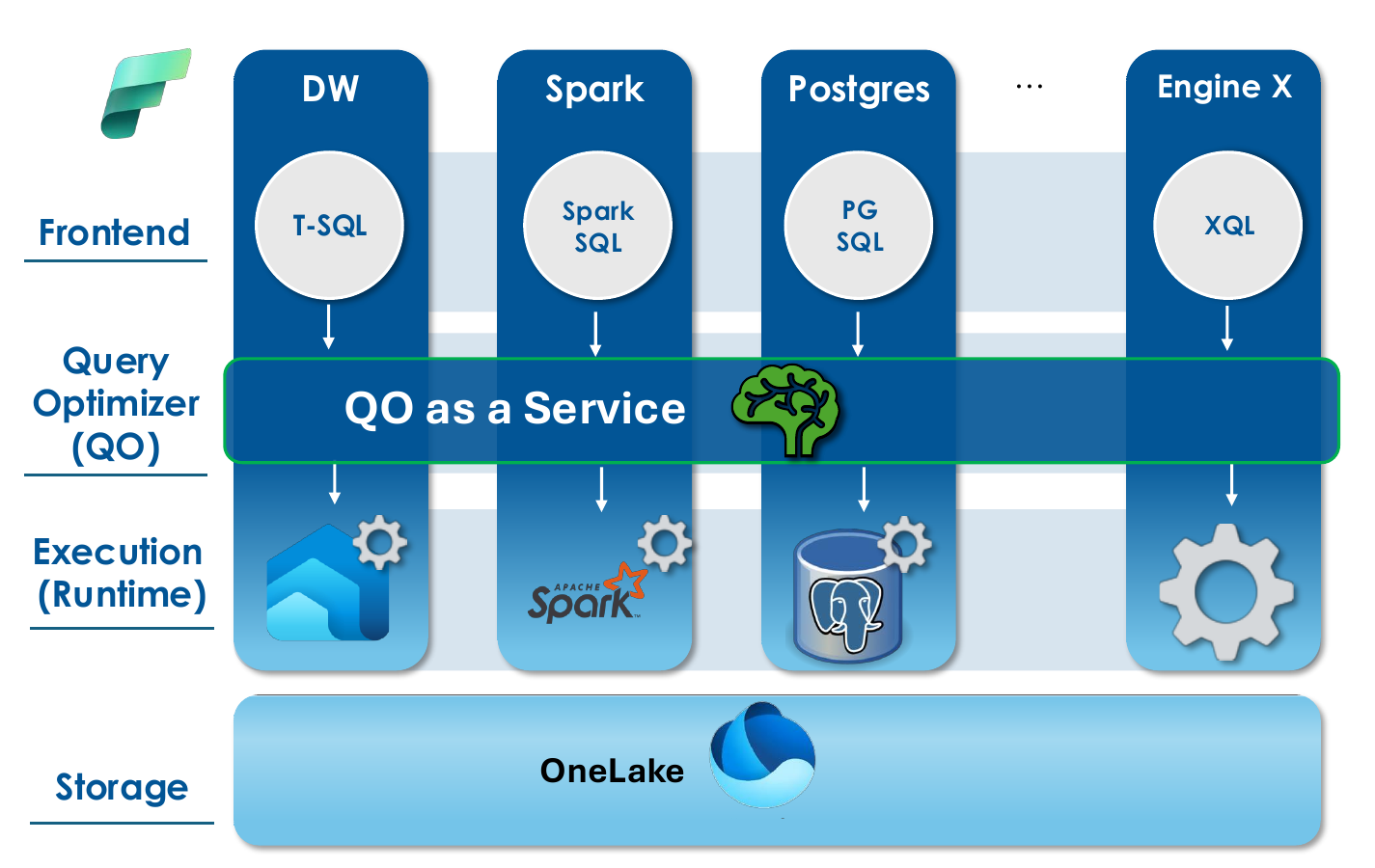}
        \vspace{-3mm}
    \caption{A unified LakeHouse ecosystem with QOaaS}
    \vspace{-3mm}
    \label{fig:qoaas}
\end{figure}

In this paper, we focus on the opportunity to unify and share components in the Query Optimization (QO) layer.
We build on our own experience in building Calcite, and in evolving the Cascades framework across SQL Server~\cite{sqlserver}, Fabric DW~\cite{FabricDW}, and SCOPE~\cite{scope} within Microsoft\footnote{The Fabric DW QO is a branch of the SQL Server QO, whereas the SCOPE QO is a fork of the SQL Server QO.}.
While these efforts provided great re-use benefits, we argue that: \emph{``It is time for something more!''}
We propose a vision for a full-blown Query Optimizer as a Service (QOaaS) in a unified LakeHouse ecosystem like Fabric, as illustrated Figure~\ref{fig:qoaas}.

QOaaS moves past the library-sharing approach of Calcite or Orca, by decoupling QO in a separate service, that interacts with potentially multiple engines over RPC. 
We base our design on a Microsoft Fabric like ecosystem, and ground our thinking with a prototype compatible with Fabric DW and Spark.
The key benefits of QOaaS (see the comparison in Table~\ref{tab:comparison_qo}) include the ability to:
1)~isolate the QO from the rest of the engine query processing, allowing for independent deployment/experimentation,
2)~centrally handle workload-level optimizations such as index/view selection and ML-enabled QO enhancements,
3)~accelerate development by amortizing costs across engines, and 4)~in a longer term, multi-engine federation, where each query sub-plan is executed by the most optimal engine.

\begin{table}
    \centering
    \footnotesize 
    \begin{tabularx}{\columnwidth}{|C||C|C|C|} \hline 
\vfill\textbf{Features} & \vfill\textbf{Custom QO} & \vfill\textbf{QO as a Library} & \vfill\textbf{QOaaS}\\ \hline
\vfill Innovation speed& \xmark & \cmark & \cmark\\ \hline 
Engineering efficiency & \xmark & \cmark & \cmark\\ \hline 
New engine time-to-market & \xmark & \cmark & \cmark\\ \hline 
Cross-engine optimization & \xmark & \xmark & \cmark\\ \hline 
\vfill QO scalability & \xmark & \xmark & \cmark\\ \hline
Workload Observability & \xmark & \xmark & \cmark\\ \hline 
Workload Optimization & \xmark & \xmark & \cmark\\ \hline 
    \end{tabularx}
    \caption{
    \revision{Comparison of QO Approaches}}
    \vspace{-7mm}
    \label{tab:comparison_qo}
\end{table}

The above gain requires us to tackle some hard technical challenges including:  
1) the need to define an \emph{unambiguous query plan format}, capable of capturing  query semantics and all their nuances\footnote{For example, PowerBI and Fabric DW \code{group-by sum()} operations have subtle semantic differences in treating categories with no entries, where PowerBI reports all keys, and where no tuples exist a 0 sum value, while Fabric DW would omit the entry if no records exist for that key.}, 
2) by unifying QO, the system must provide flexible cardinality estimation, cost models, and plan search mechanisms capable of delivering optimal plans for diverse engines\footnote{For example, DuckDB column-oriented group-by SIMD-optimized implementation will vastly differ from Spark scale-out-over-shuffle one.
}, and 
3) addressing the performance impact of interactions between different engines and the remote service for communicating all necessary metadata required for the QO to perform its work effectively.

While QOaaS is an ambitious vision, we discuss our thinking and highlight our initial stumbles and successes as we investigate the opportunity of building a commercial QOaaS. 
In full fairness, this is more than an idea but less than a committed product plan. 
We aim to use this paper and the live engagement with the savvy and opinionated CIDR audience to converge our thinking--whether to pursue it at full-speed or abandon it.

\revision{For this current QOaaS effort, we are focusing specifically on \textit{relational} \textit{analytical} engines within a \textit{unified} \textit{Lakehouse} ecosystem. First of all, as OLTP queries tend to be much simpler, analytical engines have greater QO requirements that would justify a QOaaS. Additionally, while QOaaS can work for a single data warehouse or lakehouse, it is particularly effective in a unified Lakehouse environment like Fabric, where multiple engines can operate on data in a shared lake. Although QOaaS could also potentially be applied to non-relational engines with high QO requirements, we are initially concentrating on relational engines, as implementing a QOaaS even for these use cases is already a significant challenge.}

\revision{While federated query execution is enabled by QOaaS, it is only one of the many benefits brought by QOaaS, as pointed out in Table~\ref{tab:comparison_qo}. We would like to differentiate QOaaS from the QOs of federation systems, like Garlic~\cite{garlic}, and polystore systems, like BigDAWG~\cite{elmore2015demonstration}. Both types of systems enable querying across multiple data engines, with polystores specifically targeting at heterogeneous engines with different data models. In terms of QO, which is the focus of this paper, each individual backend engine in these systems retains its own QO unchanged, functioning as a black box within the broader system. The QO component for the federated server or polystore coordinaitng across engines primarily handles slicing a query into subqueries and assigning these subqueries to the appropriate engines. In contrast, our QOaaS approach decomposes the individual engines and unifies the QO components into one shared QO, with the goal of improving engineering efficiency and innovation speed. Queries executed either by an individual engine or across engines can both benefit from QOaaS.}

\section{Reinventing wheels in the QO space}

Recent years have witnessed a proliferation of analytical query engines, each with its own QO to create efficient execution strategies. Although these QOs may appear different on the surface, with variations in the number and types of operators and transformation rules, they typically model the same relational algebra, explore similar search spaces, and follow the same stages outlined below.

\textbf{Parsing/Algebrization.} The input query is parsed into an Abstract Syntax Tree (AST) and transformed into an algebraic tree of relational operators.
Table and column names are resolved using metadata catalogs, and type inference annotates the tree while removing invalid queries.

\textbf{Simplification/Normalization.}
The input tree is iteratively transformed into an equivalent canonical form using simplification rules, applied via top-down, bottom-up, or fixed-point schemes.

\textbf{Pre-exploration.}
This phase involves tasks performed after simplifying the input tree and before cost-based exploration. Needed statistics are loaded or scheduled for creation if absent.

\textbf{Cost-based exploration/implementation.}
In this phase, physical alternatives for the input operator tree are generated, compared, and the best one is chosen.
The process involves three main pillars: the search space of alternatives, the enumeration strategy, and the cost model. 
Systems may use dynamic programming for join orders and specialized approaches for other tasks.
The Cascades Optimization Framework, used in multiple industrial optimizers \cite{sqlserver,FabricDW, scope,spanner,lyu2021greenplum,begoli2018apache,soliman2014orca}, relies on the Memo data structure for compact representation and optimization tasks for generating alternatives via transformation rules. 
\textbf{Post-optimization.}
In this phase, peephole transformations are applied to the final plan, and query execution structures are created for the selected plan.

Additional evidence that modern QOs share significant functionality comes from optimizer libraries like Calcite~\cite{begoli2018apache} and Orca~\cite{soliman2014orca}, which are reused by multiple engines.
At Microsoft, the SQL Server QO has been reused for SCOPE~\cite{scope} and Fabric DW~\cite{FabricDW} with minor extensions.
These examples also demonstrate that QOs can function as an independent building block rather than an integral part of the monolithic data engine.

\section{Steps towards QOaaS}

In this section, we describe our initial efforts to realize the QOaaS vision within a unified LakeHouse ecosystem like Fabric. 
As a testing vehicle for the feasibility of QOaaS, we decided to adapt the Unified Query Optimizer (\uqo)~\cite{bruno2024unified}, which is the QO for Fabric DW, to optimize queries for other engines, building on our experience of evolving SQL Server QO to optimize other systems. 
This decision was also influenced by our initial evaluation of both Fabric DW and Fabric Spark engines, which share significant commonalities. 
With this context, we focus on three key challenges for QOaaS using \uqo: exchanging query plans in and out of the QO (Section~\ref{subsec:substrait}), adapting the optimizer to work effectively with different engines (Section~\ref{subsec:spark_uqo}), and adjusting the cost model to accommodate for implementation differences in various engines (Section~\ref{sec:qotuning}).

\subsection{Standardizing Plan Specification}
\label{subsec:substrait}
For a QOaaS to operate across multiple engines, a unified, cross-language, and cross-engine plan specification is essential. 
As recommended in~\cite{composabledb}, we adopted Substrait~\cite{substrait} to standardize intermediate plan representation across engines within Fabric. Substrait is an open-source project that provides a standard, language-neutral specification for relational algebra, supporting various serialization formats like Protobuf and JSON. It allows extensions for custom operations and includes a robust ecosystem of libraries and tools for validation and implementation.

We have been working to replace engine-specific plan representations with Substrait across Fabric engines such as DW, Spark, and Power BI. 
Our current implementation has achieved support for: 1)~various SQL operations such as scan, filter, joins (including inner, outer, semi, and anti-semi), grouping, and sort,
2)~a wide range of expressions, including scalar (arithmetic, boolean, cast) and aggregate functions, and
3)~both simple types such as different precision integers, floating point, and date-time, as well as compound types such as decimal and fixed- and variable-length character strings.
This support is sufficient to cover queries from TPC-H and TPC-DS, as well as certain internal Microsoft workloads. 
With Substrait serving as the cross-engine plan representation, we can now mix and match different QOs and runtimes for a query.

\begin{figure}[tbh]
    \centering
    \vspace{-3mm}
    \includegraphics[width=0.38\textwidth]{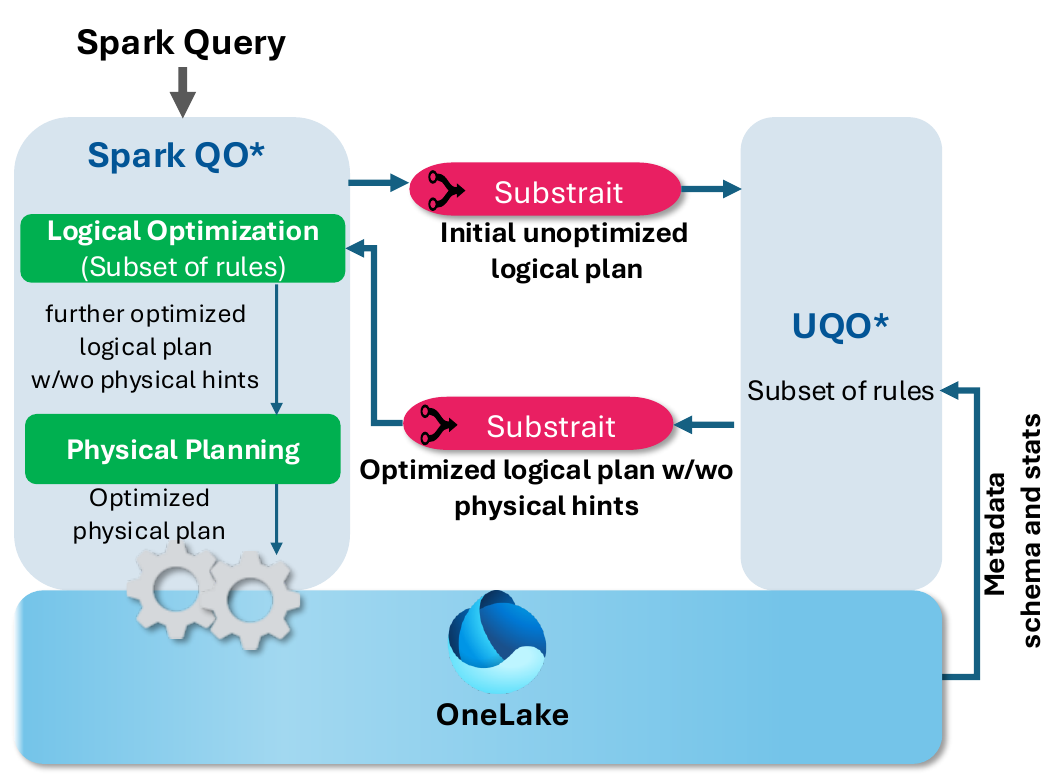}
    \vspace{-3mm}
    \caption{Optimizing Spark queries with \uqo}
    \vspace{-6mm}
        \label{fig:qo-for-spark}
\end{figure}

\subsection{Optimizing Spark Queries with \uqo}
\label{subsec:spark_uqo}

\revision{To test the concept of QOaaS, we focused on the two analytics engines within Fabric, DW and Spark. Currently, the Spark query optimizer primarily relies on non-cost-based optimizations, with limited cost-based optimization (CBO) features, such as join reordering and broadcast-vs-shuffle join decisions. We have long been wondering about the potential advantages of replacing the Spark optimizer with UQO, which encompasses 255 CBO rules. To explore this, we built a prototype of QOaaS using UQO to optimize Spark queries, as depicted in Figure~\ref{fig:qo-for-spark}, and evaluate the effectiveness of using such a QOaaS.}

To manage complexity and expedite development, we initially targeted queries that use operators and functions supported by both DW and Spark. 
To optimize Spark queries effectively with \uqo, our prototype needed to address two main categories of mismatches. 
First, there are \textbf{physical operator mismatches}.
The set of physical operators supported by DW and Spark differs, which means that some physical plans generated by \uqo may not be executable in Spark due to missing corresponding physical operators. 
For instance, since Spark currently lacks support for indexes, any physical operators that rely on indexes in DW are not supported in Spark.
To address this, we disabled optimization rules in \uqo that generate unsupported operators. \revision{We denote this modified \uqo as \uqostar in Figure~\ref{fig:qo-for-spark}}.
Second, there are \textbf{feature support mismatches}.
Some Spark features, such as Hive-style partitioning and distributed Bloom filters, were not supported in the version of \uqo we were experimenting with. 
As a result, even if \revision{\uqostar} produces a valid plan for Spark after disabling certain optimization rules, the plan might not leverage these advanced features, leading to sub-optimal performance. 
To address this issue in an initial prototype, we deliberately avoided the engineering-heavy task of extending \revision{\uqostar} to cover Spark's specific physical implementation details. 
Instead, \revision{as illustrated in Figure~\ref{fig:qo-for-spark}, we stack \uqostar with a modified Spark QO, denoted as Spark QO*, which complements \uqostar with the missing Spark optimization rules. In this prototype, a Spark query is first parsed and compiled by Spark QO*; then the unoptimized logical plan is handed over to \uqostar via the Substrait standard specification, and optimized by \uqostar to produce an optimized logical plan either with or without physical hints (more details below); subsequently, this plan is passed back to Spark QO*, via the Substrait specification, and re-optimized, with limited logical optimization and physical planning, to produce the final physical plan that can be executed in Spark.}

This QOaaS prototype implements two alternative approaches for optimizing Spark queries. 
1)~\textbf{QOaaS-v1:} \revision{\uqostar} generates optimized logical plans, and \revision{Spark QO*} further refines these plans and selects the physical implementation. 
2)~\textbf{QOaaS-v2:} \revision{\uqostar} produces optimized logical plans with hints on some physical implementation choices (e.g. choosing broadcast vs. shuffle or hash join vs. sort merge join). 
These hints are embedded in the Substrait plan using its extensibility system. 
\revision{Spark QO*} is then modified to consider these hints and generate the physical plans suggested by \revision{\uqostar}.

\begin{figure}[htbp]
    \centering
    \vspace{-4mm}
    \includegraphics[width=0.43\textwidth]{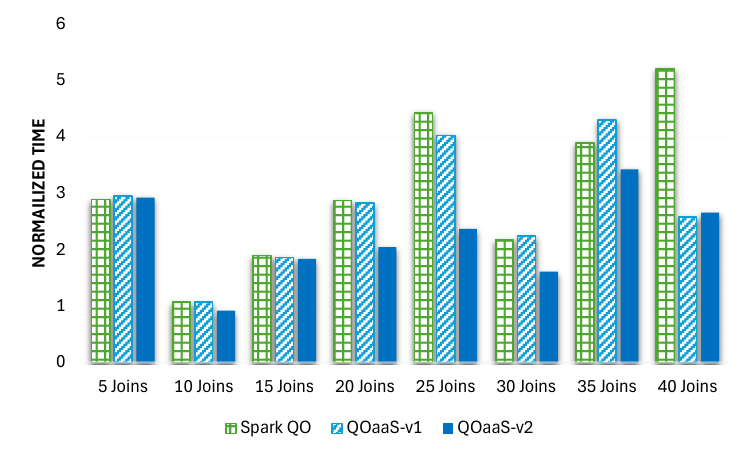}
    \vspace{-4mm}
    \caption{Execution time of representative MSSales queries on Spark runtime with different QOs}
    \vspace{-6mm}
        \label{fig:qo-for-spark-mssales}
\end{figure}

\myparagraph{Evaluation.}
We tested both QOaaS approaches with a real analytical production workload from Microsoft’s internal revenue reporting platform, MSSales, on a Spark Fabric cluster. 
This workload consists of 627 tables stored as Parquet files in OneLake, totaling around 5TB. 
The queries contains regular patterns (i.e., many queries share the same template with different parameter values) and are particularly join-heavy. 
We selected 8 representative queries with varying numbers of joins for the experiment. 

As shown in Figure~\ref{fig:qo-for-spark-mssales}, QOaaS-v1, where \uqo only contributes logical optimization, generally provides performance comparable to Spark QO. 
However, for 40-way joins, it  improves performance by nearly 2$\times$. 
This improvement is primarily because \uqo produces a better join ordering than Spark QO. 
With additional hints on the physical implementation, QOaaS-v2 significantly improves performance for most queries further. 
The more sophisticated cost-based optimization in \uqo often produces better physical plans overall. 
Even with this simple QOaaS prototype, we already observe great potential. 
Another advantage of bringing \uqo to Spark is its built-in support for view matching. This automatically enables materialized view (MV) support for Spark, allowing Spark queries to use MVs created by DW, a feature currently absent in Spark.

We also tested QOaaS-v2 on TPC-H SF1000 (1TB). 
The queries in TPC-H are less complex than those in MSSales, and Fabric Spark has already been highly tuned for this benchmark workload.
As a result, the runtime with QOaaS-v2 plans is very comparable to that with Spark QO, with an average difference within 6\%. 
However, for Query~5, we observed a 1.5$\times$ slowdown with QOaaS-v2. 
This slowdown occurs because the optimization decisions made by \uqo before Spark's optimization prevent Spark QO from applying the Bloom filters optimization effectively. 
This indicates that adding optimizations retroactively is not ideal, as noted by previous works~\cite{bruno2024unified}. 
Considering all optimizations in \uqo's search space exploration and cost decisions jointly would lead to a better plan.

\subsection{Recalibrating and Tuning the Cost Model}\label{sec:qotuning}

Like most major QOs in use today~\cite{sqlserver,FabricDW, scope,lyu2021greenplum,begoli2018apache,soliman2014orca}, \uqo employs a formula-based cost model, where the total cost is formulated as a weighted sum of the costs of basic operations on the estimated data items. 
Two integral parts of the cost formula are the \textit{estimated cardinality}, which represents the number of rows processed by each operation, and the \textit{cost parameters}, which capture the relative impact of hardware and software factors in the cost model, such as sequential disk IO, random disk IO, memory usage of a hash table, CPU cost of evaluating a predicate, etc. 
Like many other QOs, the cost parameters in \uqo use predefined, hard-coded values. 

In the context of QOaaS, it is unrealistic to expect that the same hard-coded parameter values will always produce optimal plans across engines, given the different architecture and implementation details. 
In fact, even for Fabric DW, depending on the hardware and software configuration, these cost parameters should be recalibrated (e.g. the relative impact of CPU and IO costs will differ across hardware generations and SKUs). 

As a result, we focused next on automatically tuning the cost parameters for the \uqo cost model leveraging MLOS~\cite{mlos_github, MLOS},  
a general-purpose, ML-powered infrastructure and methodology for continuous, robust, and trackable systems optimization. 
It has been applied to auto-tune many systems within Microsoft~\cite{MLOS-demo}.

We started by tuning 60 cost parameters in \uqo for a single node backend in DW. 
To help reduce the search space, we utilized LlamaTune~\cite{llamatune} to reduce the search dimension to 10. 
We also experimented with various tuning optimizers in MLOS (e.g., FLAML \cite{wang2021flaml}, SMAC \cite{lindauer2022smac3}, GridSearch) and settled on FLAML. 

\begin{figure}[htbp]
    \centering
    \vspace{-2mm}
    \includegraphics[width=0.35\textwidth]{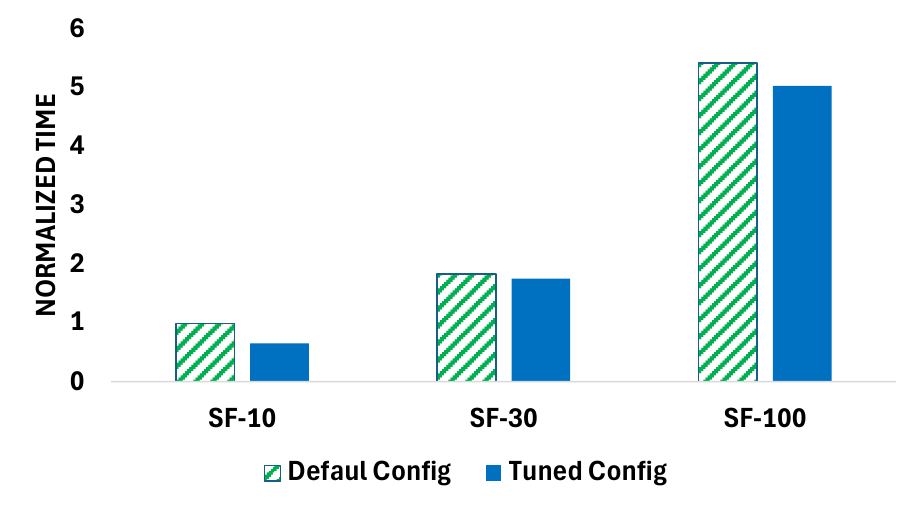}
    \vspace{-3mm}
    \caption{Runtime performance with default vs tuned parameters on TPC-H for three scale factors}
    \vspace{-4mm}
        \label{fig:qotune-tpch}
\end{figure}

\begin{figure}[htbp]
    \centering
     \vspace{-3mm}
    \includegraphics[width=0.45\textwidth]{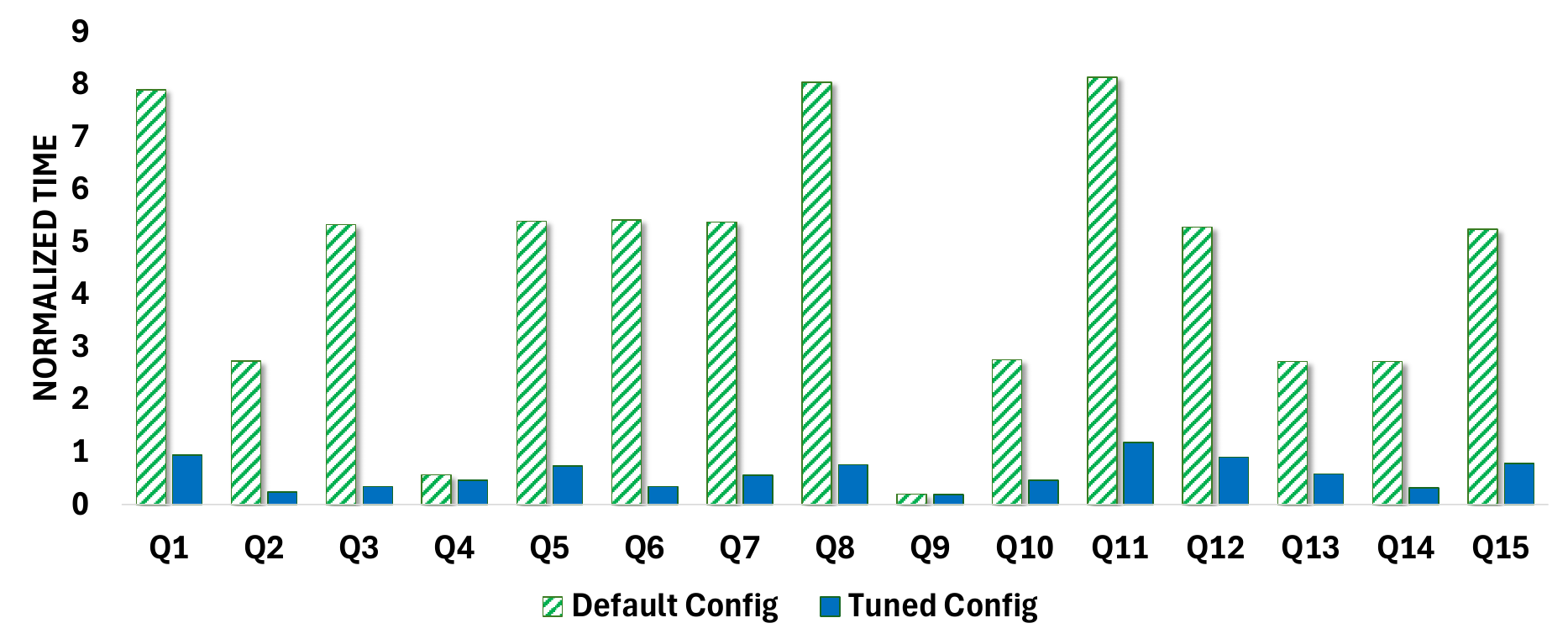}
    \vspace{-3mm}
    \caption{Runtime performance with default vs tuned parameters on a subset of MSSales data} 
    \vspace{-4mm}
        \label{fig:qotune-mssales}
\end{figure}

Figure~\ref{fig:qotune-tpch} shows the total runtime of TPC-H queries for three scale factors with the default and MLOS-tuned parameters. 
TPC-H has also been heavily tuned for DW, so the performance improvement with the tuned parameters is not very significant. 
In contrast, we witnessed substantial performance improvements, up to 8$\times$, for the MSSales workload, shown in Figure~\ref{fig:qotune-mssales}. 
Note that we used a 10GB subset of the MSSales data to keep the tuning process manageable. 

These preliminary results demonstrate that recalibrating the cost parameters for the given hardware is necessary to produce good query plans. 
However, the tuned parameters are not \textit{transferable} across workloads. 
Running the MSSales workload with the tuned parameters from TPC-H SF10 yields almost the same runtime performance as the default parameters. 
Conversely, running TPC-H SF10 with the tuned parameters from MSSales results in a dramatic performance degradation, up to 7$\times$. 
Several factors might contribute to this problem. 
First, MSSales queries contain many joins and unions but do not cover all operators in DW, causing the tuned parameters to overfit for a subset of operators and not work well for TPC-H queries containing other operator types.
Second, cardinality estimation plays an important role in the cost model; with the large number of joins in MSSales, the tuned parameters might over-compensate for cardinality estimation errors. 

Considering these observations, we have started working on two concurrent efforts: 
1)~designing a benchmark workload with good coverage of all operators and their various code paths, and 
2)~injecting true cardinalities into the cost model, leveraging the work in~\cite{ce}, to mitigate the problem of cardinality estimation errors in the tuning process. 
In addition, we plan to utilize our QOaaS prototype introduced in Section~\ref{subsec:spark_uqo} to tune \uqo cost parameters for the Spark runtime. 

It is also worth noting that even if we fail to develop a universal set of parameters that performs well across all workloads, tuning for specific workloads 
may still be valuable. In fact, we should allow instance-based optimization, like the many learning-based QO works~\cite{learnedCE,cardlearner,costmodel, learnedqo, knobadvisor, knobadvisor1}, in the design of QOaaS.

\subsection{Key Lessons Learned}

We have gleaned important insights from our initial attempt:

\begin{itemize}
    \item A standard plan specification is essential for QOaaS.
    \item The ``patchy'' approach of combining two QOs is suboptimal for QOaaS, as demonstrated in Section~\ref{subsec:spark_uqo}. Instead, QOaaS should explore all possible optimization opportunities to produce an optimal plan. 
    \item The cost model of a QOaaS must consider the engine capabilities and generate engine-specific costs accordingly.
    \item A QOaaS should allow instance-based optimization, embodied by the recent advancements in ML-based QO research.
    \item Fiddling with a production-level customized QO, like \uqo, for QOaaS requires significant engineering effort, even for simple proof-of-concept prototypes.
\end{itemize}

These insights have prompted us to rethink and propose a new design for QOaaS, which will be discussed in the next section.

\section{The QOaaS Proposal}

Building on our initial experiments and preliminary results, we believe QOaaS is a promising idea. 
In this section, we sketch out an architectural proposal for realizing the QOaaS vision. 

\begin{figure}[htbp]
    \centering
    \vspace{-3mm}
    \includegraphics[width=0.43\textwidth]{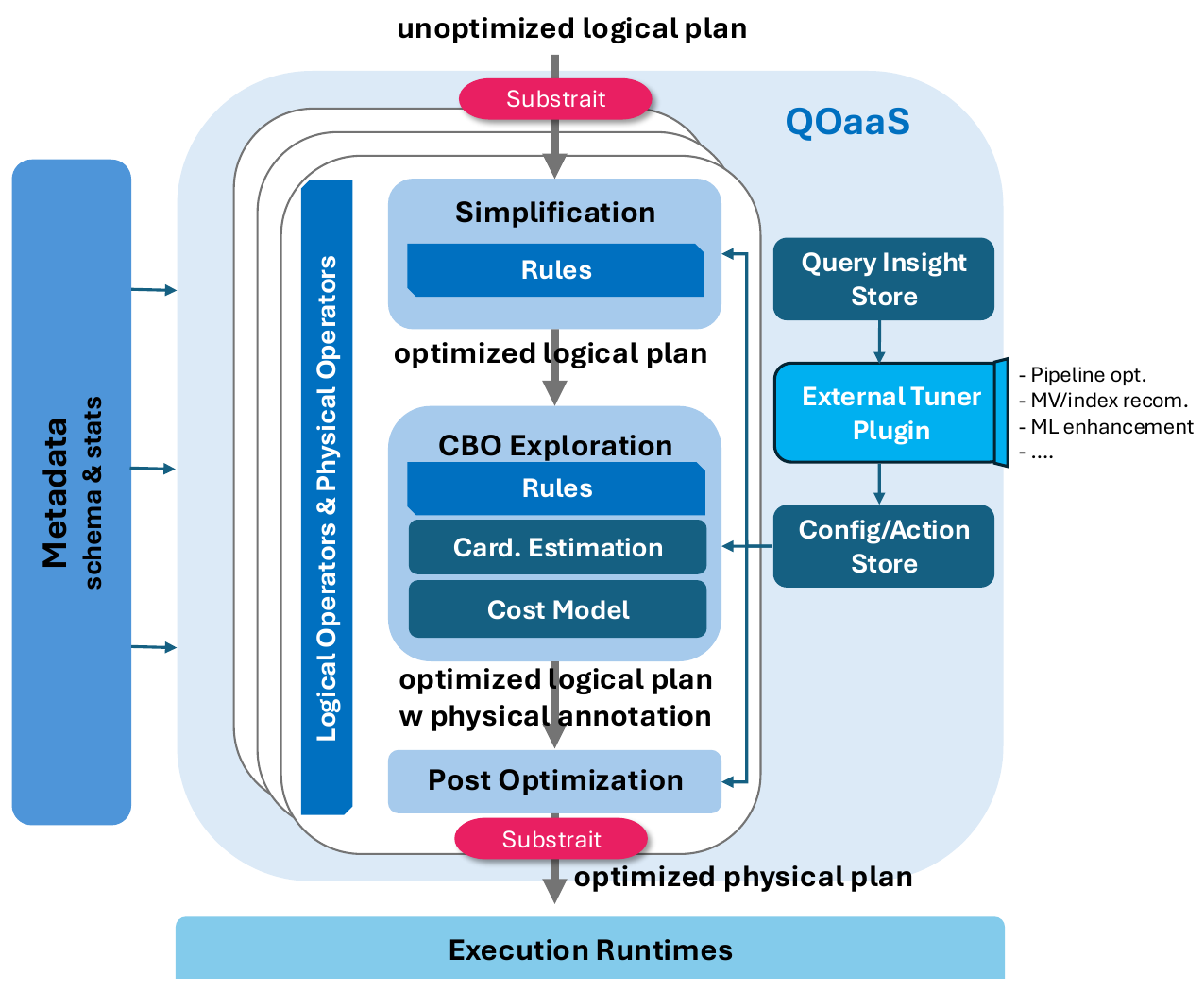}
    \vspace{-3mm}
    \caption{Proposed QOaaS architecture } 
    \vspace{-6mm}
        \label{fig:qoaas-blueprint}
\end{figure}

\subsection{Overview} 

The high-level goals for QOaaS include: serving multiple engines based on their capabilities, ease of extension to support new engines, the capability to optimize across engines even with hybrid or federated plans, and the flexibility to incorporate the latest advanced QO features, including various ML-based QO enhancements~\cite{learnedCE,cardlearner,costmodel, learnedqo, knobadvisor, knobadvisor1} that learn from historical data and proactively improve query performance.

Figure~\ref{fig:qoaas-blueprint} shows the proposed QOaaS architecture, striving to meet these goals. 
First, QOaaS adopts a standard plan specification, like Substrait~\cite{substrait}, to ensure cross-engine operability. 
At its core, there are three phases of optimization: \textit{Simplification}, which applies rewrite rules without a cost model; \textit{CBO Exploration}, which explores the plan search space with the help of a cost model; and \textit{Post Optimization}, which applies engine-specific customizations. 
We modularize each fundamental QO building blocks and make each QO component extensible, similar to the general-purpose QO libraries like Calcite and Orca. 
However, we enhance extensibility by introducing three additional components into QOaaS: a \textit{Query Insight Store}, an \textit{External Tuner Plugin} module, and a \textit{Config/Action Store}. 
These components collect past query plans and runtime statistics, allow external processes to use them for tuning, and then feed the tuning configurations or actions back to the core QO components for advanced QO enhancements.

Finally, the servicification of a unified QO allows us to elastically scale up and out (with multiple instances) the QO independently from the runtime engines, ensuring the efficiency and resilience of the optimization process with dedicated resources. 
In fact, individual components within QOaaS can also be made into standalone services. To further ensure efficiency, caching the metadata accessed by the QO becomes critical, as numerous network round trips and large payloads can cause significant overheads. 
The evaluation and evolution of the QO also become easier with the ability to perform A/B testing and run different versions of QO instances, enabling faster and easier innovation, development, and testing. 

\revision{

Note that QOaaS also necessitates a unified metadata representation across engines. 
Lakehouses already leverage open table formats such as Delta Lake~\cite{delta-lake}, Apache Hudi~\cite{apache-hudi}, and Apache Iceberg~\cite{apache-iceberg}, along with interoperability tools like Apache XTable~\cite{xtable}, to unify the management of both data and metadata versioning, as well as other essential functionalities within data lakes. 
Complementing this, file formats like Puffin~\cite{apache-iceberg-puffin} introduce mechanisms for storing versioned statistics through flexible, arbitrary blobs. 
Engines are not \emph{required} to interpret or update these blobs today, but there is clear potential to standardize blob specifications (e.g., histograms, sketches) and enable multi-language bindings via formats like Protobuf, facilitating access from multiple engines. 
Building on these advancements, recent proposals like OpenHouse~\cite{openhouse}, Apache Polaris~\cite{apachepolaris}, and Unity Catalog~\cite{unity} focus on unified operational metadata catalogs to manage table definitions, schemas, and governance. 
While formalizing such standards is beyond the scope of this paper, these efforts reflect the broader momentum toward unified metadata in the Lakehouse ecosystem.
}

\subsection{Core Components}

\textbf{Operators and Rules.} Operators (logical or physical) and rules are the fundamental building blocks of a QO.
While many operators and rules are commonly shared across different engines, some are specific to certain engines. 
To make QOaaS aware of which operators and rules are supported by each engine, we add a \textit{engines-property} to each operator and rule, indicating in which engine(s) it is supported. 
We also add new exchange operators that support cross-engine data exchange (including transfer and transformation if needed). 
Similar to any other physical property like sort order, the engines-property will be enforced during optimization.

\textbf{Simplification Phase.}
This phase applies a series of simple logical transformations to the unoptimized logical plan, like constant folding and filter pushdown. 
Based on the engines-property, QOaaS only applies those rules supported by the target engine(s).

\textbf{CBO Exploration Phase.}
In this phase, QOaaS applies cost-based optimizations using two modularized sub-components: \textit{Cost Model} (CM) and \textit{Cardinality Estimation} (CE). 
During the CBO exploration, based on the engines-property, QOaaS only applies those rules supported by the target engine(s). 
Besides the normal input for a cost model, the CM component in QOaaS also takes a target engine as an additional input. 
Depending on the target engines, it might employ different cost models. 
For instance, it might use different cost parameters for Spark compared to those for DW while employing the same cost formula, or it could utilize an entirely different cost formula, or even a non-formula-based cost model. 
If a plan needs to be optimized across multiple engines, costs for different engines for the same (sub-)plan are computed and compared during exploration, and exchange operators might be enforced based on the engines-property. 
This phase employs both logical and physical optimization rules. 
However, since we do not want to burden the CBO with too much physical implementation details of each engine (e.g. the whole-stage code generation in Spark), this phase only produces optimized logical plan with all necessary physical annotation. 
The physical annotation includes physical operators (e.g. hash join vs sort merge join) and the chosen target engines.

\textbf{Post Optimization Phase.}
We handle all engine-specific implementation details in this phase. 
Logical plans with physical annotations are transformed into physical plans, incorporating all necessary implementation details, including the chosen target engine for each operator. 
If a hybrid plan is produced, the physical plan includes all necessary information to execute the hybrid plan.

\subsection{New Components}

The core components of the QOaaS architecture share resemblance with the modularized components in a general QO library like Calcite or Orca, but the three new components, Query Insight Store, External Tuner Plugin, and Config/Action Store, are unique for QOaaS.
They enhance the existing QO with observability across engines and open it up for advanced new augmentations.

The introduction of these components draws inspiration from past efforts in extending the SCOPE QO (also based on SQL Server QO) with a service-oriented architecture~\cite{oasis} for computation reuse~\cite{cloudview} and learned QO components~\cite{knobadvisor1, knobadvisor, cardlearner, costmodel}, collaboratively conducted by the Gray Systems Lab and the SCOPE team.

\textbf{Query Insight Store.}
All learning-based QO improvements~\cite{learnedCE,cardlearner,costmodel, learnedqo, knobadvisor, knobadvisor1} require training data to begin with. 
In order to allow novel ML-based QO components pluggable into the QOaaS, we need to first enable observability of past workloads. 
The Query Insight Store is introduced to automatically capture a history of queries, plans, and runtime statistics (such as execution time and actual cardinality) across engines in the unified LakeHouse. 
These data are collected from the query logs, QOaaS itself, and the telemetries emitted by the various engines in the LakeHouse.

\textbf{External Tuner Plugin.}
This component is the door that opens QOaaS for trusted external processes or services which can leverage the information in Query Insight Store for QO enhancement and then store the new configuration or action for the QOaaS in the Config/Action Store.
This component offers a set of APIs to access data in the Query Insight Store using both pull and push mechanisms, with optional filters for retrieving specific subsets of observables.
Additionally, it offers APIs for storing information into the Config/Action Store.

\textbf{Config/Action Store.}
The new configurations or actions generated by an external tuner process or service are stored in the Config/Action Store, so that the core components of the QOaaS can access them and utilize them for better optimization.

\subsubsection{Examples of Extensions to QOaaS}
We now provide some examples of utilizing the three new components to extend QOaaS with more advanced features.

\textbf{Example 1: Cost model parameter tuning.}
To enable the cost model parameter tuning (see Section~\ref{sec:qotuning}), we can plugin an external process through External Tuner Plugin, which accesses the past queries in the Query Insight Store and selects a set of representative queries for each target engine. 
Then it calls MLOS to conduct offline tuning for each engine with the selected workload. 
The final tuned cost parameters are then stored into the Config/Action Store as updated defaults.
Finally, the Cost Model component in QOaaS can be extended to use these different cost parameters for each engine.

\textbf{Example 2: Learned cardinality estimator.}
A QOaaS that observes all engines in a unified LakeHouse gives us an advantage of leveraging the rich statistics collected from multiple engines to train a better ML-based cardinality estimator. 
We can use a pluggable external ML training process to build models using the plans and true cardinalities from the Query Insight Store. 
The learned models can be stored back into the Config/Action Store, which are finally used by an extended Cardinality Estimation module. This process can be run periodically to update the models with new data.

\textbf{Example 3: Pipeline optimization.} 
A QOaaS offers a unique opportunity for pipeline optimization, particularly when the pipeline involves jobs across different engines.
By observing all queries in a pipeline, QOaaS can conduct a global analysis to optimize the entire pipeline, similar to~\cite{pipemizer}. 
For instance, an upstream query in Spark might produce data without specific ordering or partitioning, while multiple downstream jobs in DW could benefit from a particular order or partitioning scheme.
In this case, QOaaS can recommend adjustments to the Config/Action Store, enabling the optimizer to automatically add a sort or partition operator to the producer job.
\vspace{-2mm}
\subsection{Challenges and Risks}

A coin has two sides. 
Naturally, the proposed QOaaS architecture comes with its own set of challenges and risks. 

\textbf{QO complexity:} While customizability and extensibility are generally advantages for QOaaS, they also increase the software complexity of the QO.

\textbf{Learning curve:}
Consolidating the QO developers into one team is beneficial, but it requires the QO developers to understand the overall requirements of multiple engines.

\textbf{Coordination across teams:}
The cross-engine optimization capability of QOaaS requires close coordination among the relevant teams.
This includes agreeing on data collection for the Query Insight Store, establishing protocols for using information in the Config/Action Store, and designing any new APIs and extensions.

\textbf{Communication overheads:}
Separating the QO into a remote service can increase communication costs from metadata transfer between engines and the QO.
Factors like caching and session maintenance also come into play. 
Central metadata services can reduce these overheads but require further coordination of standards. 

\textbf{Innovation hurdle:}
While QOaaS can accelerate innovation by allowing the core components to be shared across multiple engines--meaning a single change can benefit multiple engines--it also presents a challenge. 
Changes that improve one engine could negatively impact others. 
As a result, rolling out updates requires more thoughtful design and rigorous testing to ensure compatibility and stability across engines.
However, the potential for easy A/B testing provides an opportunity to mitigate this issue by allowing for careful validation before widespread deployment.

\section{Open Discussion and Debate}

Current industry trends have pointed us to a future with a more unified and composable data management architecture. 
In this spirit, we conjectured an ambitious vision of a QOaaS that rules all the query engines in a converged LakeHouse ecosystem like Microsoft Fabric. 
\textit{Is this a good idea?} 
\textit{Will it work?} 
We do not have a definitive answer yet. 
But we started with concrete baby steps towards this grand vision. 
Based on trial and error with some real production systems on Fabric and past successes from related efforts, we have sketched out a preliminary design for QOaaS. 
The goal of this design is not to provide a solution but rather to spark a conversation and debate. 
Perhaps the idea of \textit{one QO to rule them all} is a fantasy. 
However, maybe a QOaaS for a few very similar engines (e.g. Fabric DW and Spark) is both feasible and beneficial. 
We invite the CIDR community to critique and discuss with us on this exciting topic.

\bibliographystyle{ACM-Reference-Format-num}
\bibliography{reference}

\end{document}